\documentclass[twocolumn]{aastex631}

\usepackage{pinlabel}
\usepackage{amsmath}

\begin{document}

%\title{Template \aastex Article with Examples: 
%v6.3.1\footnote{Released on March, 1st, 2021}}

\title{Improved Carbon and Nitrogen Isotopic Ratios for CH$_3$CN in Titan's Atmosphere Using ALMA}

\accepted{PSJ, March 2025}

%correspondingauthor{August Muench}
%email{greg.schwarz@aas.org, gus.muench@aas.org}

\correspondingauthor{Jonathon Nosowitz}
\email{nosowitz@cua.edu}

\author[0009-0004-1366-9472]{Jonathon Nosowitz}
\affiliation{Department of Physics, The Catholic University of America, Washington, DC 20064, USA}
\affiliation{Solar System Exploration Division, NASA Goddard Space Flight Center, Greenbelt, MD 20771, USA}

\author[0000-0001-8233-2436]{Martin A. Cordiner}
\affiliation{Solar System Exploration Division, NASA Goddard Space Flight Center, Greenbelt, MD 20771, USA}
\affiliation{Department of Physics, The Catholic University of America, Washington, DC 20064, USA}

\author[0000-0001-9540-9121]{Conor A. Nixon}
\affiliation{Solar System Exploration Division, NASA Goddard Space Flight Center, Greenbelt, MD 20771, USA}

\author[0000-0002-8178-1042]{Alexander E. Thelen}
\affiliation{Division of Geological and Planetary Sciences, California Institute of Technology, Pasadena, CA 91125, USA}

\author[0000-0002-2570-3154]{Zbigniew Kisiel}
\affiliation{Institute of Physics, Polish Academy of Sciences, Al. Lotników 32/46, 02-668 Warszawa, Poland}

\author[0000-0003-3108-5775]{Nicholas A. Teanby}
\affiliation{School of Earth Sciences, University of Bristol, Bristol BS8 1RJ, UK}

\author[0000-0002-6772-384X]{Patrick G. J. Irwin}
\affiliation{Atmospheric, Oceanic and Planetary Physics, Clarendon Laboratory, University of Oxford, Oxford OX1 3PU, UK}

\author[0000-0001-6752-5109]{Steven B. Charnley}
\affiliation{Solar System Exploration Division, NASA Goddard Space Flight Center, Greenbelt, MD 20771, USA}

\author[0000-0001-7273-1898]{V\'{e}ronique Vuitton}
\affiliation{Univ. Grenoble Alpes, CNRS, IPAG, 38000 Grenoble, France}



%250 word limit
\begin{abstract}

Titan, Saturn's largest satellite, maintains an atmosphere composed primarily of nitrogen (N$_2$) and methane (CH$_4$) that leads to a complex organic chemistry. Some of the nitriles (CN-bearing organics) on Titan are known to have substantially enhanced $^{15}$N abundances compared to Earth and to Titan's dominant nitrogen (N$_2$) reservoir. The $^{14}$N/$^{15}$N isotopic ratio in Titan's nitriles can provide better constraints on the synthesis of nitrogen-bearing organics in planetary atmospheres as well as insights into the origin of Titan's large nitrogen abundance. Using high signal-to-noise ratio ($>13$), disk-integrated observations obtained with the Atacama Large Millimeter/submillimeter Array (ALMA) Band 6 receiver (211-275 GHz), we measure the $^{14}$N/$^{15}$N and $^{12}$C/$^{13}$C isotopic ratios of acetonitrile (CH$_3$CN) in Titan's stratosphere. Using the Nonlinear optimal Estimator for MultivariatE spectral analySIS (NEMESIS), we derived the CH$_3$CN/$^{13}$CH$_3$CN ratio to be 89.2 $\pm$ 7.0 and the CH$_3$CN/CH$_3$$^{13}$CN ratio to be 91.2 $\pm$ 6.0, in agreement with the $^{12}$C/$^{13}$C ratio in Titan's methane, and other Solar System species. We found the $^{14}$N/$^{15}$N isotopic ratio to be 68.9 $\pm$ 4.2, consistent with previously derived values for HCN and HC$_3$N, confirming an enhanced $^{15}$N abundance in Titan's nitriles compared with the bulk atmospheric N$_2$ value of $^{14}$N/$^{15}$N = 168, in agreement with chemical models incorporating isotope-selective photodissociation of N$_2$ at high altitudes.    

\end{abstract}



\section{Introduction} \label{sec:intro}

The origin and evolution of the atmosphere of Titan, Saturn's largest satellite, has been an area of interest for many years. Titan's atmosphere is substantial, with intricate chemistry for a moon, and can provide a template for understanding the compositions of primitive carbon/nitrogen-dominated (exo-)planetary atmospheres. As we continue to investigate Titan's atmosphere, more questions arise on topics such as cloud and haze formation, chemical pathways, interactions with surface and sub-surface features, and the full extent of Titan's organic inventory \citep{nixon_2018, mackenzie_2021}. Titan possesses a thick atmosphere composed primarily of molecular nitrogen (N$_2$) at about 98\% and methane (CH$_4$) at about 2\% in the stratosphere, but the origins of these surprisingly abundant gases are not well known. Titan's methane was first observed by \citet{kuiper_1944}, and various ideas have been put forward to explain its presence and persistence \citep{nixon_2012}, such as diffusion from subsurface reservoirs \citep{kossacki_lorenz_1996}, episodic outgassing from clathrate hydrates \citep{tobie_2006, choukroun_2010}, or release from cryo-lava flows \citep{davies_2016}.

\citet{glein_2015} suggested that the gases responsible for forming Titan's dense, nitrogen-rich atmosphere were originally trapped in its core, and that subsequent hydrothermal and cryovolcanic processes were critical to the formation of Titan's atmosphere. However, this is reliant on chemical reactions, outgassing, and transport mechanisms to produce the gas abundances currently observed. \citet{glein_2015} showed that this method is plausible through mass balance and chemical equilibrium calculations, but also acknowledged that missing information, such as a value for a reasonable outgassing efficiency, makes it challenging to conclusively validate the theory. Thus, an external (cometary) source of Titan's nitrogen may be plausible. The isotopic ratios derived for cometary gases are similar to those of Titan's atmosphere, suggesting that the primordial NH$_3$ reservoir represented by cometary ices could be the source of Titan's nitrogen \citep{mandt_2014}.         

The N$_2$ and CH$_4$ in Titan's atmosphere form the basis of a complex chemical reaction network, yet the chemistry involving nitrogen is still not fully constrained \citep{horst_2017, nixon_2024}. It is particularly important to understand the chemistry of nitriles, as they are the most abundant nitrogen-containing photochemical products and may play a role in prebiotic syntheses \citep{oro_1990}, in addition to their importance in the organic chemistry of space, and the potential for life beyond Earth in general. 

Nitriles often possess a large dipole moment, and thus their rotational transitions can be detected in Titan's atmosphere at mm/sub-mm wavelengths, including HCN, HC$_3$N, CH$_3$C$_3$N, C$_2$H$_5$CN and more \citep{horst_2017, cordiner_2015, palmer_2017, thelen_2020, marten_2002, rengel_2011, rengel_2022}. These molecules, and a diverse population of other organics, are generated in Titan's upper atmosphere through high-altitude photochemistry, following dissociation by UV photons, and collisions with charged particles from Saturn's magnetosphere, and/or galactic cosmic rays. 

Previously developed photochemical models (\textit{e.g.} \citealt{dobri_loison_2018, vuitton_2019, wilson_atreya_2004, willacy_2016}) obtain a moderately good agreement with the observed abundances of many nitrogen-bearing species in Titan's atmosphere, suggesting good progress in our quantitative understanding of the relevant chemical processes. However, there still exist significant gaps in our knowledge of Titan's photochemistry, in terms of the detailed reaction pathways as a function of altitude. The available photochemical models make different assumptions regarding the relevant reaction pathways, and the $^{14}$N/$^{15}$N ratios in Titan's nitriles are particularly sensitive to some of these assumptions. Enrichment (or isotopic fractionation) of $^{15}$N relative to Titan's bulk N$_2$ reservoir is theorized to occur as a result of isotope selective photodissociation of N$_2$ \citep{liang_2007}, which involves the more efficient self-shielding of $^{14}$N$_2$ compared with the less abundant $^{14}$N$^{15}$N isotopologue at high altitudes, leading to a reservoir of gas-phase atomic nitrogen that is isotopically enriched in $^{15}$N. The resulting $^{15}$N enrichment is readily passed on to nitrogen-bearing photochemical products. However, the atomic $^{14}$N/$^{15}$N ratio is observationally unconstrained, and is sensitive to the various model parameters. Incorporation of $^{15}$N into nitriles occurs at different rates depending on the altitude-dependent reaction pathways; indeed, the $^{14}$N/$^{15}$N isotopic ratio in CH$_3$CN has been shown to depend on the relative efficiencies of the different formation pathways \citep{dobri_loison_2018}, which are not yet fully constrained by experiments. Further studies of the $^{14}$N/$^{15}$N ratios in Titan's nitriles are therefore needed, to improve our understanding of the relevant chemical processes, which will lead to a better understanding of nitrogen chemistry (including isotope chemistry) in planetary atmospheres.

The first $^{14}$N/$^{15}$N isotopic ratio measurement of Titan's atmosphere was by \citet{marten_2002}, who used the Institut de Radioastronomie Millim\'etrique (IRAM) 30-m telescope to derive a value of HC$^{14}$N/HC$^{15}$N = 60--70 in the stratosphere. This was followed by \citet{gurwell_2004} using the Submillimeter Array, who obtained  94$\pm$13 and 72$\pm$9 depending on the assumed temperature profile. \citet{vinatier_2007} subsequently derived HC$^{14}$N/HCN$^{15}$N = 56 $\pm$ 8 using Cassini infrared spectroscopy. \citet{courtin_2011} used the Spectral and Photometric Imaging Receiver (SPIRE) instrument on the Herschel spacecraft to derive a ratio of 76 $\pm$ 6. The HC$^{14}$N/HC$^{15}$N ratio was further refined by \citet{molter_2016} who obtained 72.2 $\pm$ 2.2 using high spectral-resolution ALMA observations. Using in-situ mass spectrometry, Cassini-Huygens measured the atmospheric $^{14}$N$_2$/$^{15}$N$^{14}$N ratio, from which a $^{14}$N/$^{15}$N ratio of $167.7\pm0.6$ was derived in N$_2$, which represents the dominant atmospheric nitrogen reservoir \citep{niemann_2010}. 

More recently, \citet{cordiner_2018} used ALMA to obtain a $^{14}$N/$^{15}$N ratio of $67\pm14$ in HC$_3$N, while \citet{palmer_2017}, and \citet{iino_2020} measured CH$_3$C$^{14}$N/CH$_3$C$^{15}$N ratios of 89$\pm$5, and $125^{+145}_{-44}$, respectively. While there is some apparent scatter among the various measurements, a significant difference is evident between the $^{15}$N fractions in the (trace) photochemical products and the (bulk) nitrogen (N$_2$) reservoir, which can be explained as a result of isotope selective N$_2$ photodissociation by solar radiation \citep{liang_2007,vuitton_2019}. The significance of any differences in the degree of $^{15}$N enrichment among different nitriles is yet to be investigated in detail. As one of the most abundant nitriles in Titan's atmosphere, more accurate measurements of the CH$_3$CN isotopologues, in particular, are justified, to help test and improve models for Titan's nitrogen chemistry.

\section{Observations} \label{sec:observations}

To investigate Titan's high-altitude nitrogen chemistry, we obtained observations using the Atacama Large Millimeter/submillimeter Array (ALMA) Band 6 receiver (211-275 GHz; $\sim$1.1-1.4 mm), in 2019 as part of project \#2019.1.00783.S. Three spectral settings were observed (see \cite{thelen_2020} for details), the first of which (Science Goal 1; SG1) covered the CO $J=2-1$ line, and the remaining two (SGs 2 and 3) covered observations of multiple nitrile species and their carbon and nitrogen isotopes. The data from SGs 2 and 3 used in the present study were observed in several non-consecutive frequency intervals between 256.9--257.8~GHz and 267.0--268.9~GHz (see Table \ref{tab:data_table}).

As described by \citet{thelen_2020}, the data were taken during multiple execution blocks between November 14 and December 16, 2019 using ALMA configurations C43-1, C43-2 (maximum baselines ranging from 160 to 314 m), and C43-3 (maximum baselines of 500 m). The resulting beam size was 1.54$''\times$1.14$''$ in the 256-257 GHz range and 1.56$''\times$1.11$''$ in the 267-268 GHz range, so Titan ($\approx1.0''$ diameter on the sky, including its solid disk and extended atmosphere) was not resolved, enabling the maximum sensitivity per beam for disk-averaged studies of Titan's entire Earth-facing atmosphere. The spectral resolution of the CH$_3$CN and CH$_3$$^{13}$CN isotopologues was 488~kHz and the resolution for the $^{13}$CH$_3$CN and CH$_3$C$^{15}$N isotopologues was 976~kHz. The data were processed and calibrated using version 5.6.1-8 of the Common Astronomy Software Applications (CASA) pipeline using standard scripts provided by the Joint ALMA Observatory (JAO). Additional bandpass calibration smoothing was performed, to improve the spectral noise per channel, and the {\fontfamily{qcr} \selectfont tclean} procedure was used to reconstruct the sky model. For a more complete description of the observations and data processing, we refer the reader to \citet{thelen_2020}. 

A list of the observed CH$_3$CN spectral lines of relevance to the present study, their frequencies and upper state energies ($E_u$) are shown in Table \ref{tab:data_table}. The total signal-to-noise ratios (SNRs) integrated across the full-width of all detected CH$_3$CN lines are: 1450 for the major isotopologue, 13 for $^{13}$CH$_3$CN, 15 for CH$_3$$^{13}$CN and 16 for CH$_3$C$^{15}$N.

\section{Radiative Transfer Modeling} \label{sec:rad_trans_modeling}

From the processed data cubes, disk-averaged spectra were extracted using a circular aperture with a radius of 1.8$''$, which includes Titan up to the top of its atmosphere plus half a beam to account for all emission. Titan's disk was divided into 35 annuli corresponding to different emission angles from 3-75 degrees covering the center of Titan up to a tangential altitude of 1200 km \citep{teanby_2013}; the vertical extent of our model. Spectral models were generated using the line-by-line module of the Nonlinear optimal Estimator for MultivariatE spectral analySIS (NEMESIS) radiative transfer and retrieval tool \citep{irwin_2008}. NEMESIS applies an iterative minimization technique to the cost function (including the goodness of fit, $\chi^2$, combined with the a-priori errors) in order to obtain an optimized spectral model, temperature, and abundance profiles as a function of altitude. NEMESIS includes the necessary physical parameters such as spontaneous emission and absorption of radiation, on and off-limb emission angles, continuum opacity and thermal and pressure broadening, as well as temperature dependence when computing model fluxes as a function of wavelength. This allows a full characterization of the observed spectral line profiles, leading to robust abundance measurements.  

\begin{table}[]
    \centering
    \caption{Observed CH$_3$CN isotopologue transitions}
    \begin{tabular}{c c c c c c c}
        \hline
        \hline
        Species & Transition\footnote{Rotational transitions are denoted as ${J^\prime}_{K^\prime}-J^{\prime\prime}_{K^{\prime\prime}}$ where the transition is from the upper state to the lower state and K represents the angular momentum quantum number} & Rest. Freq. & $E_u$ \\
                &   & (GHz)       & (K) \\ [0.5ex]

        \hline
        CH$_3$CN & 14$_6$-13$_6$ & 257.3491793 & 349.7 \\
        CH$_3$CN & 14$_5$-13$_5$ & 257.4035843 & 271.2 \\
        CH$_3$CN & 14$_4$-13$_4$ & 257.4481277 & 206.9 \\
        CH$_3$CN & 14$_3$-13$_3$ & 257.4827915 & 156.9 \\
        CH$_3$CN & 14$_2$-13$_2$ & 257.5075614 & 121.2 \\
        CH$_3$CN & 14$_1$-13$_1$ & 257.5224275 & 99.8 \\
        CH$_3$CN & 14$_0$-13$_0$ & 257.5273835 & 92.7 \\

        \hline
        CH$_3$$^{13}$CN & 14$_3$-13$_3$ & 257.3555752 & 156.9 \\
        CH$_3$$^{13}$CN & 14$_2$-13$_2$ & 257.3802430 & 121.2 \\
        CH$_3$$^{13}$CN & 14$_1$-13$_1$ & 257.3950476 & 99.8 \\
        CH$_3$$^{13}$CN & 14$_0$-13$_0$ & 257.3999832 & 92.6 \\

        \hline 
        CH$_3$C$^{15}$N &  15$_5$-14$_5$ & 267.4868839 & 281.6 \\
        CH$_3$C$^{15}$N &  15$_4$-14$_4$ & 267.5323198 & 217.2 \\
        CH$_3$C$^{15}$N &  15$_3$-14$_3$ & 267.5676777 & 167.1 \\
        CH$_3$C$^{15}$N &  15$_2$-14$_2$ & 267.5929435 & 131.3 \\
        CH$_3$C$^{15}$N &  15$_1$-14$_1$ & 267.6081070 & 109.9 \\
        CH$_3$C$^{15}$N &  15$_0$-14$_0$ & 267.6131621 & 102.7 \\

        \hline 
        $^{13}$CH$_3$CN & 15$_5$-14$_5$ & 267.8245852 & 281.7 \\
        $^{13}$CH$_3$CN & 15$_4$-14$_4$ & 267.8698337 & 217.3 \\
        $^{13}$CH$_3$CN & 15$_3$-14$_3$ & 267.9050464 & 167.2 \\
        $^{13}$CH$_3$CN & 15$_2$-14$_2$ & 267.9302086 & 131.5 \\
        $^{13}$CH$_3$CN & 15$_1$-14$_1$ & 267.9453102 & 110.0 \\
        $^{13}$CH$_3$CN & 15$_0$-14$_0$ & 267.9503447 & 102.8 \\
        \hline
    \end{tabular}

    \label{tab:data_table}
\end{table}

To further improve the accuracy of our model retrieval results, updates were made to the spectral line database and instrumental broadening function. The NEMESIS line database was updated to the GEISA 211 format \citep{geisa_2003} for additional precision on the rest frequencies and intensities. We initially used the line data from The Cologne Database for Molecular Spectroscopy (CDMS) \citep{muller_2001} but found discrepancies in the line intensities and partition function coefficients for the major isotopologue and $^{13}$C minor isotopologues. These values were not sufficiently accurate in this frequency range and therefore, all of the values were re-computed without hyperfine or vibrational corrections in a self-consistent manner. Note that the hyperfine splittings for all relevant transitions are much smaller than the resolution of our observed spectra. We refer the reader to Appendix \ref{sec:app_recalc_vals} for more details. The partition functions for all of the isotopologues were tabulated for inclusion in NEMESIS using a third order polynomial fit to the re-computed values listed in Table \ref{tab:partition_fcn_vals}. We also updated the instrumental line shape used by NEMESIS, to account for the intrinsic (non-Gaussian) shape of the ALMA line spread function, as well as including the potential effects of Doppler broadening of the lines due to Titan's zonal winds (see Appendix \ref{sec:app_lsf} for details). 

Optimization of the vertical abundance profile began with a reasonable guess based on prior measurements. For the major CH$_3$CN isotopologue, we used an a priori profile based on the disk-averaged measurements from \citet{marten_2002} up to 500 km, supplemented by data from the model of \cite{loison_2015} up to 1200 km. The error on this `a-priori' profile was set to a constant value of 100\%. From the N$_2$ broadening parameters of CH$_3$CN calculated by \citet{dudaryonok_2015}, we assumed an average value for the Lorentzian half-width using the $K=0$ \& $K=5$, $J=14-13$ coefficients ($\Gamma$ = 0.158 cm$^{-1}$ atm$^{-1}$) and temperature exponent ($\alpha$ = 0.60). The collision-induced absorption (CIA) parameters used include all combinations of N$_2$, CH$_4$, and H$_2$ and are calculated by following the routines of \citet{borysow_1986a, borysow_1986b, borysow_1986c, borysow_1987} and \citet{borysow_1993}. The atmospheric temperature vertical profile was obtained by modeling the ALMA CO data obtained contemporaneously with our observations (see \citet{thelen_2020} for details).  

\noindent A correlation length of 3.0 atmospheric scale heights was used to internally smooth the optimized vertical profiles. The observed ALMA flux spectra were scaled by a constant factor ($\approx1.02$) to produce a match between the modeled and observed continuum levels. This scaling mitigates errors on the ALMA flux scale (which may be as much as $\approx10$\%, due to uncertainty in the amplitude calibrator flux), as well as accounting for small errors in the assumed (a priori) temperature around the tropopause, which determines the model continuum level. The CH$_3$CN major isotopologue profile was continuously retrieved at each altitude to obtain an optimized fit to the observed spectral line profiles. The retrieved profile was found to be consistent across several forms of a priori profiles (see Appendix \ref{sec:a-priori sensitivity}) as shown in Figure \ref{fig:profile_sensitivity}.

\begin{figure*}
\centering
    \includegraphics[width=0.27\textwidth]{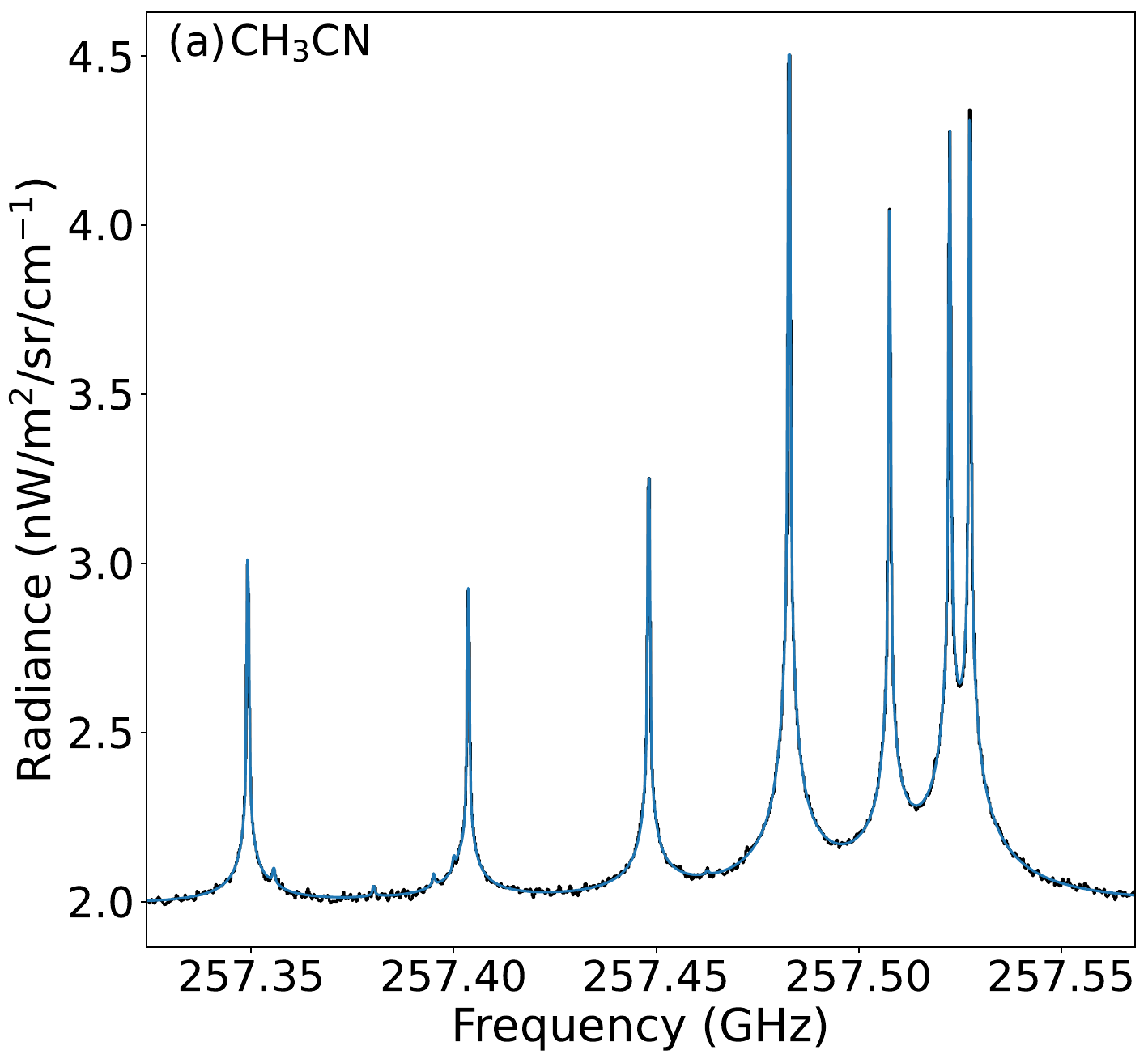}
    \includegraphics[width=0.27\textwidth]{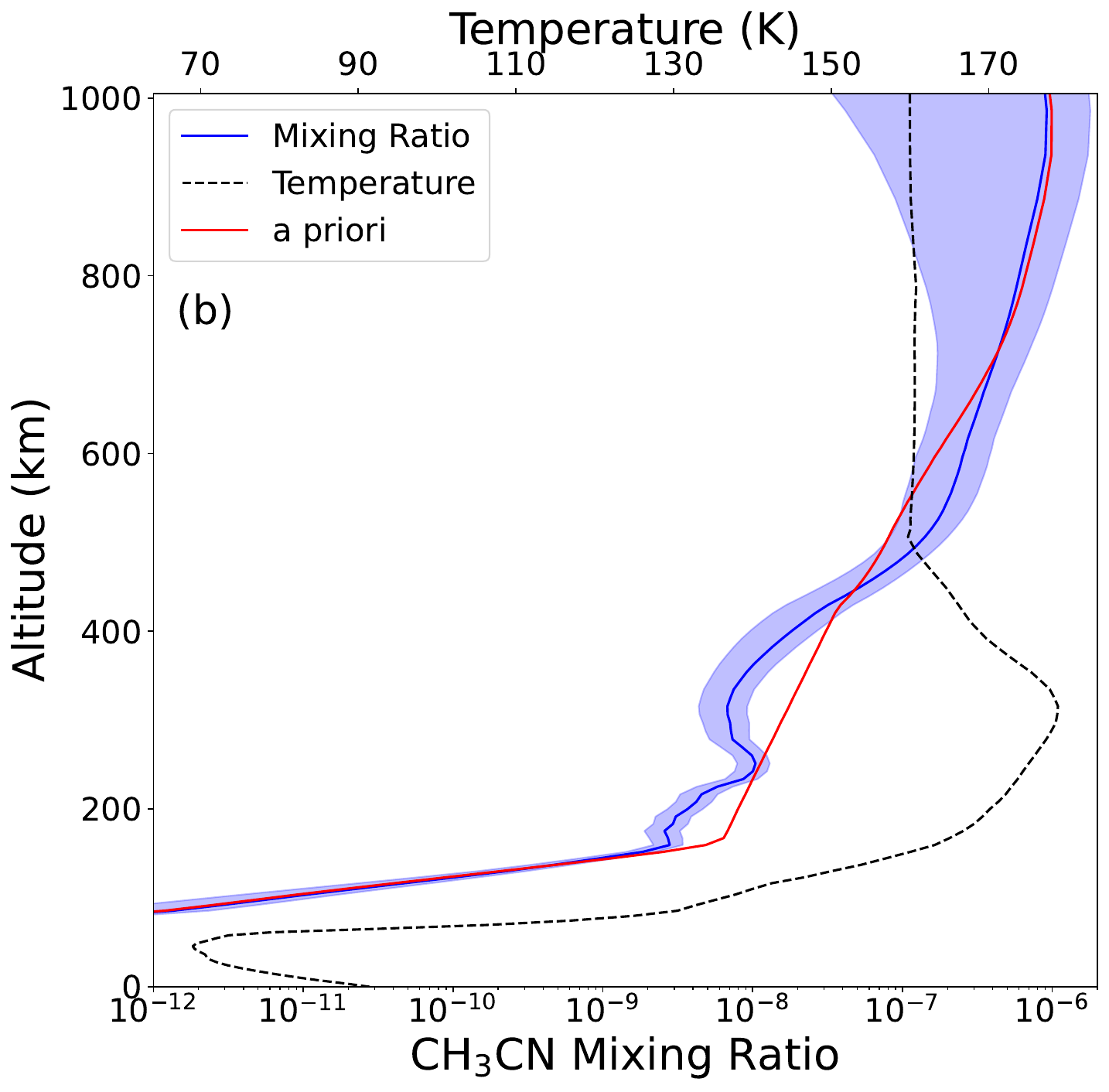}\\[3mm]
    \includegraphics[width=0.27\textwidth]{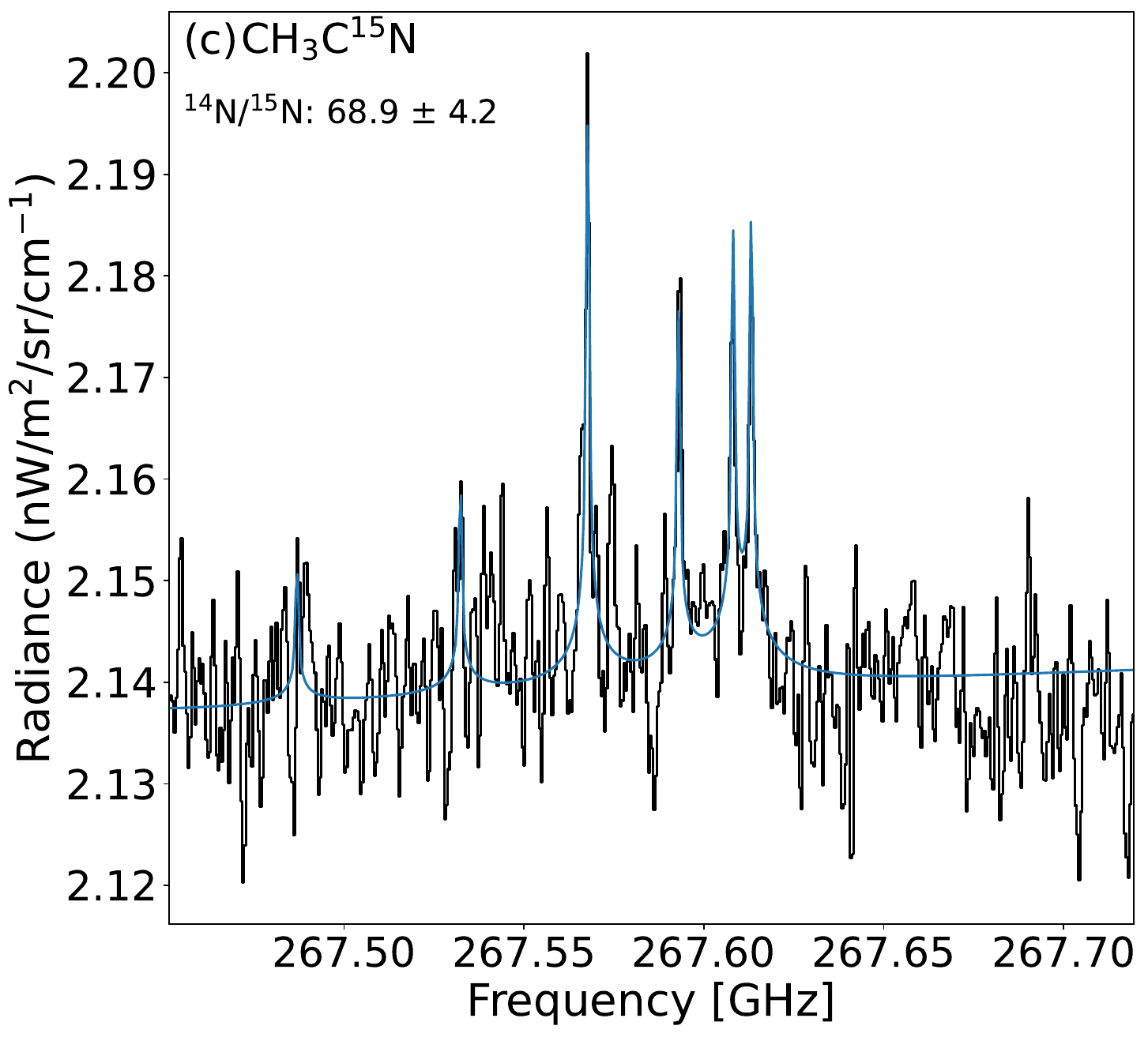}
    \includegraphics[width=0.27\textwidth]{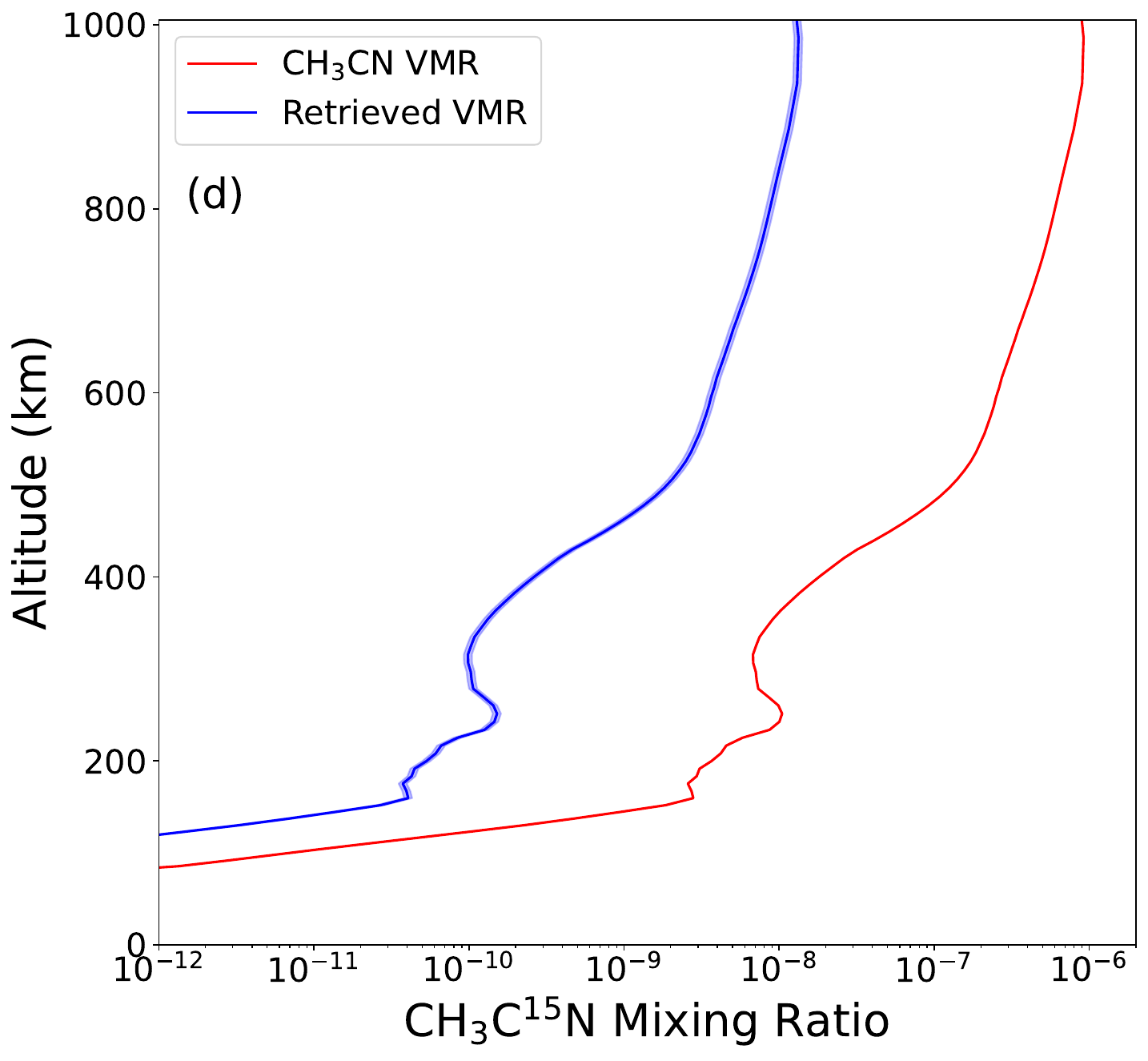}\\[3mm]
    \includegraphics[width=0.27\textwidth]{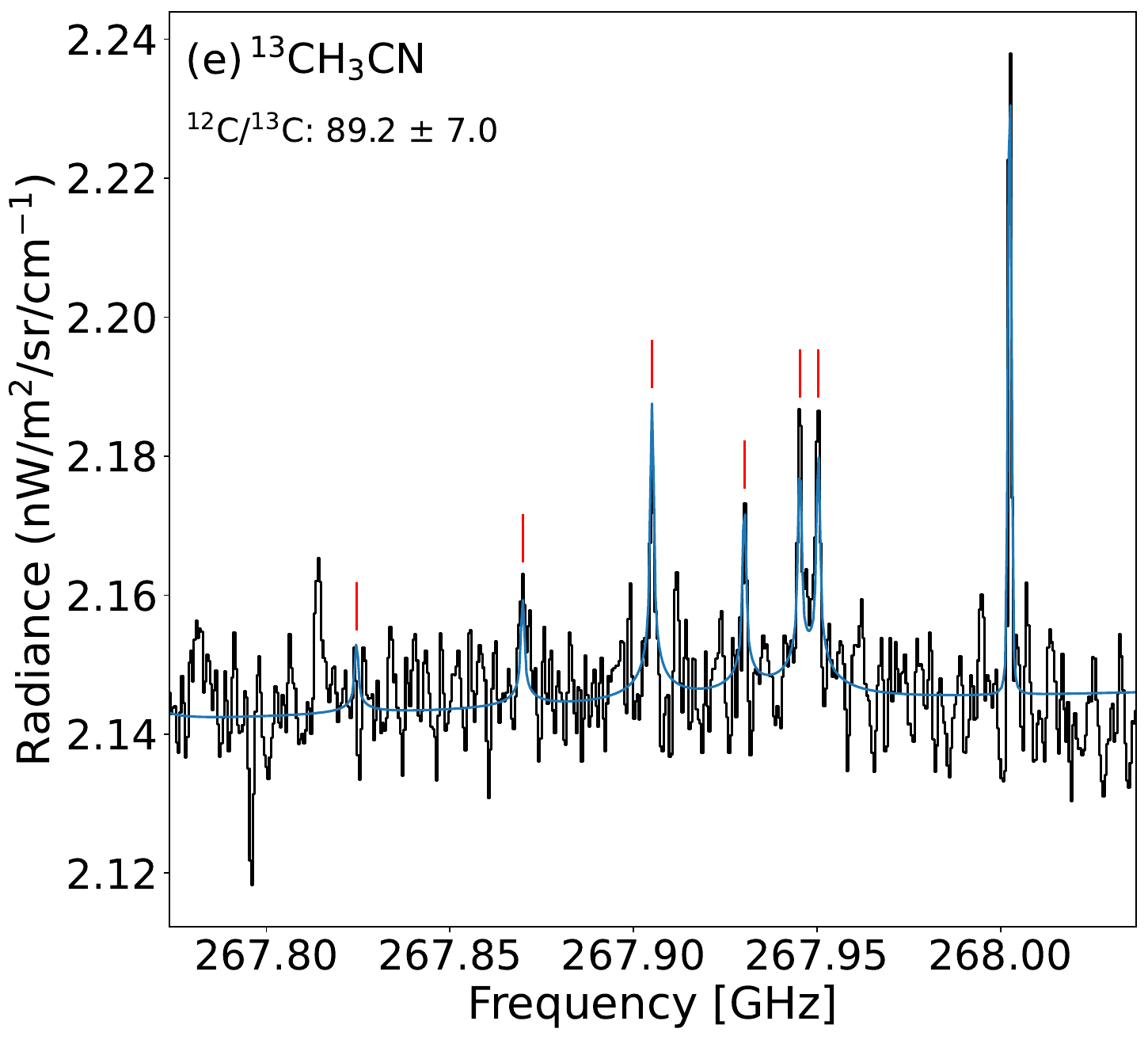}
    \includegraphics[width=0.27\textwidth]{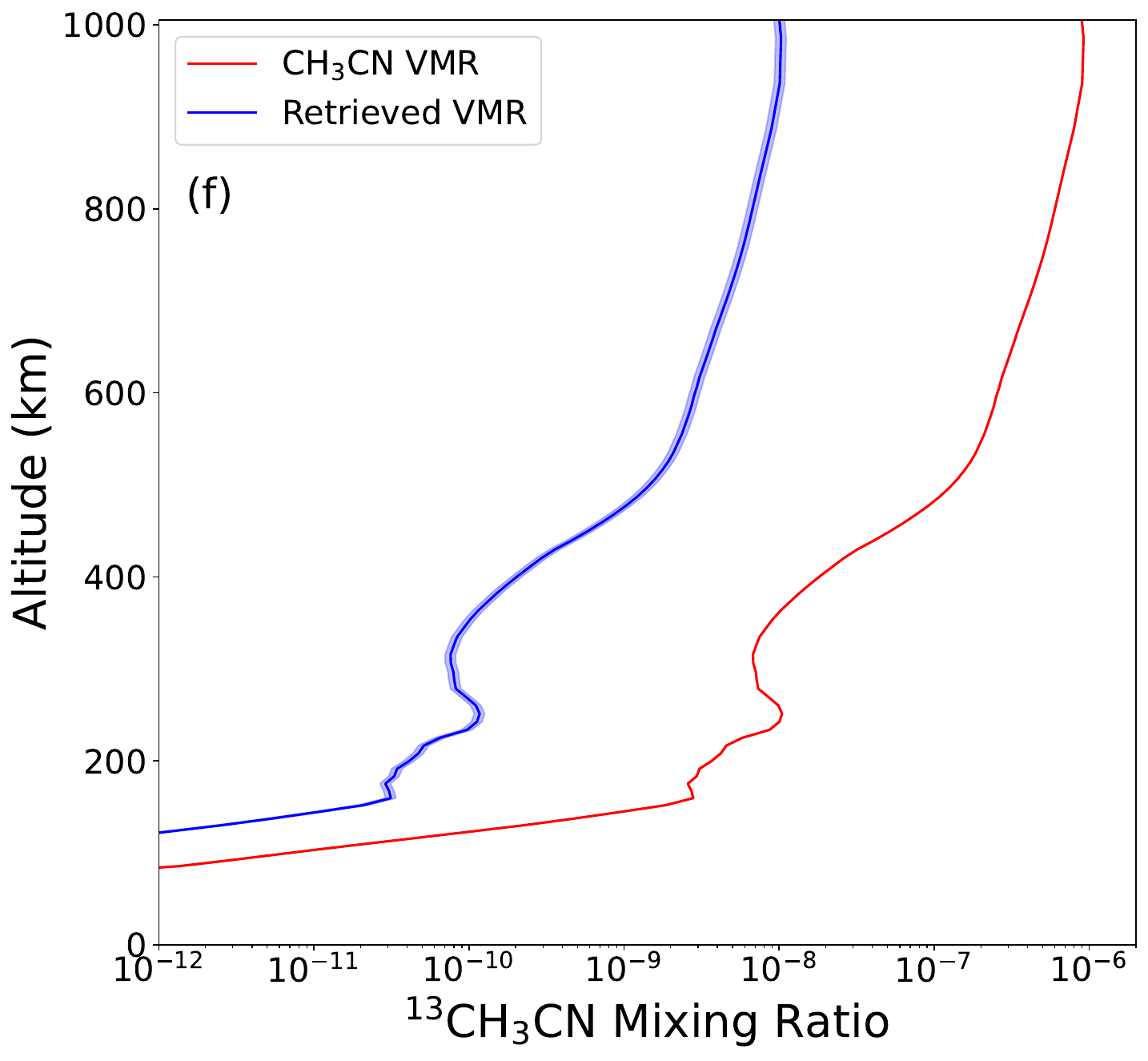}\\[3mm]
    \includegraphics[width=0.27\textwidth]{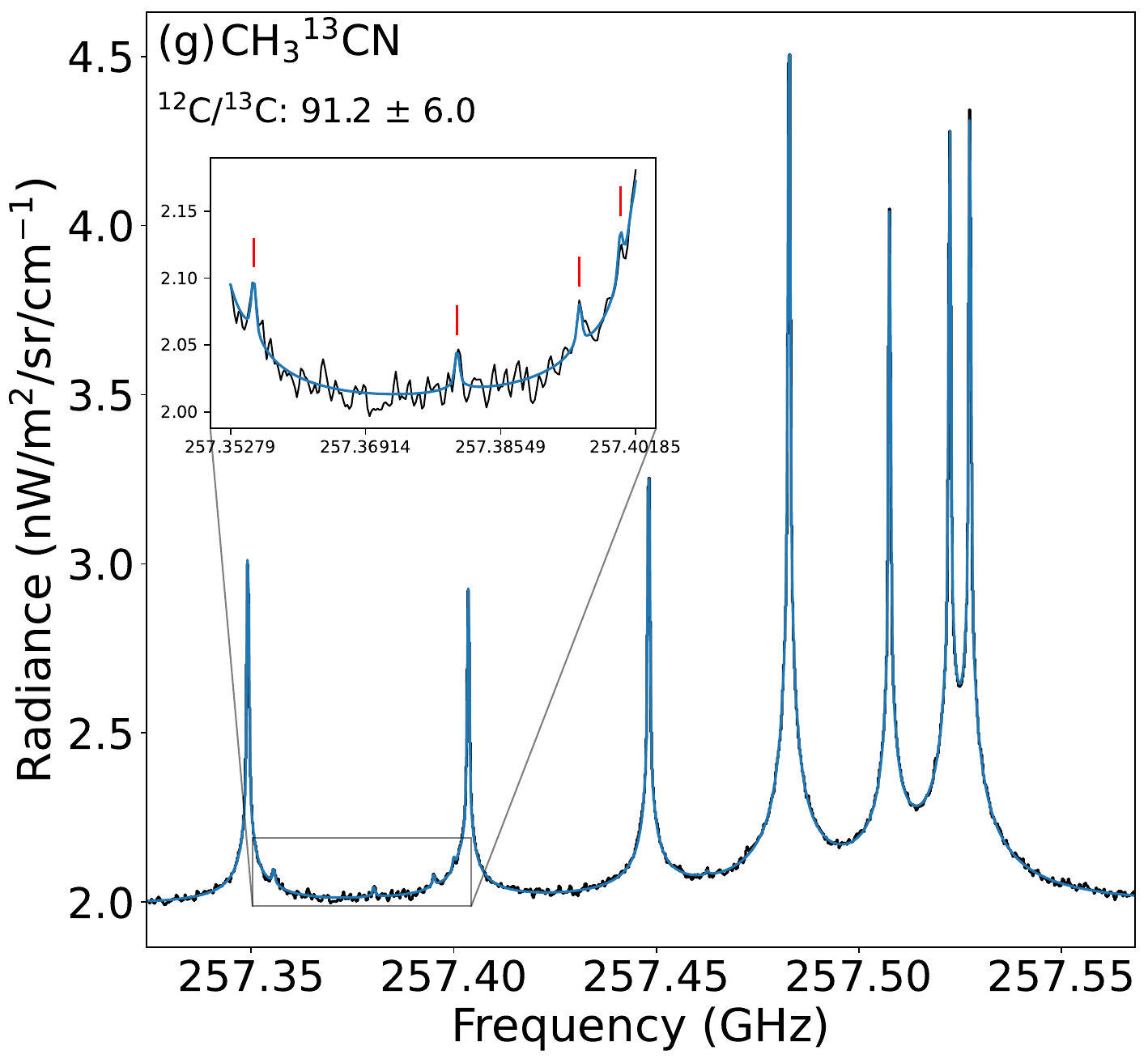}
    \includegraphics[width=0.27\textwidth]{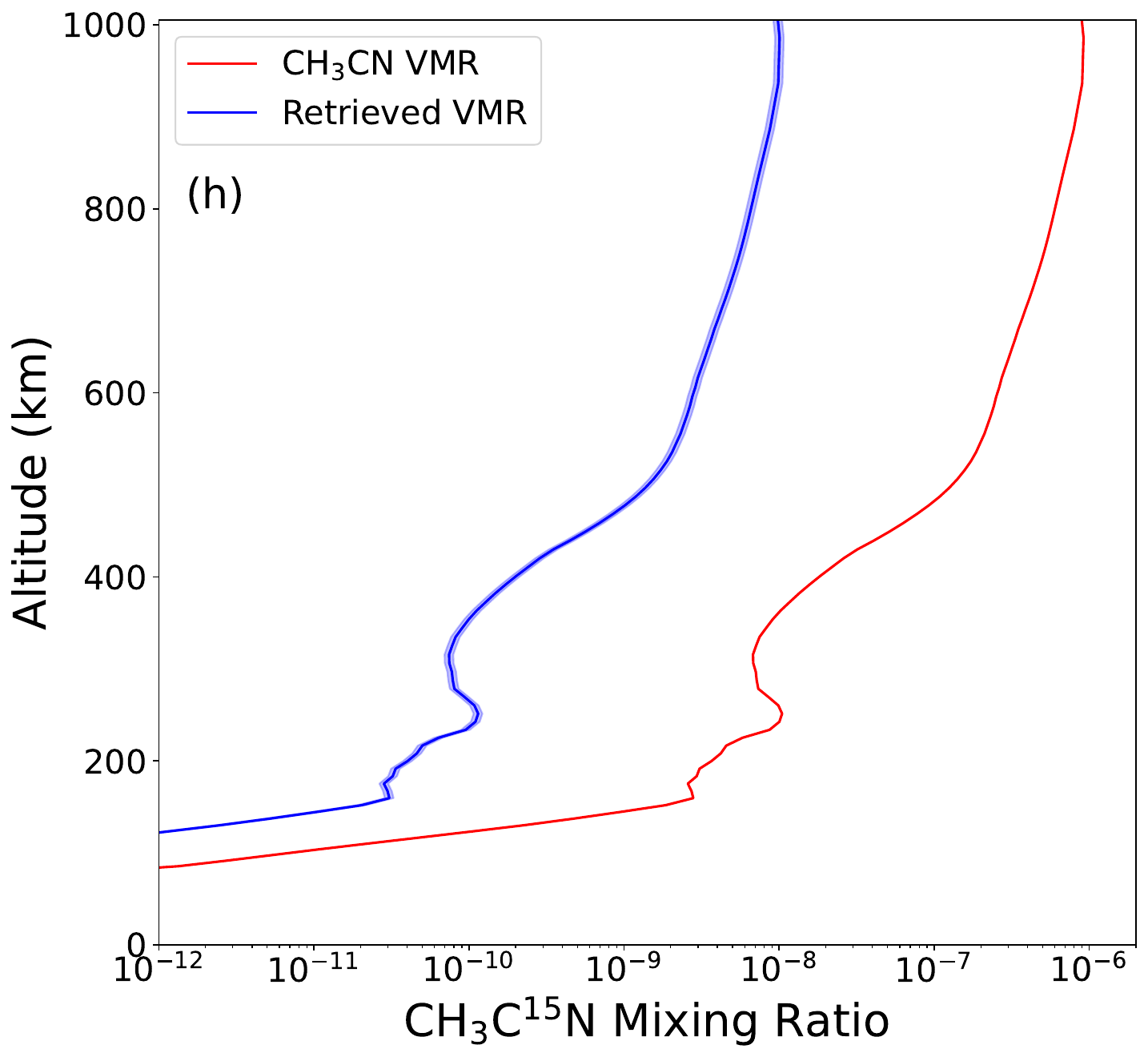}
    \caption{Optimized model fits for all isotopologues in the left column, panels (a), (c), (e), and (g). The data is in black and the model is in blue. The molecule is identified in the top left corner of the panel along with the corresponding scaling factor. The respective retrieved vertical profiles (volume mixing ratios or VMRs) are in the right column, panels (b), (d), (f), and (h). In all profiles, the red line is the a priori profile, the blue line is the retrieved VMR profile and the shaded region is the error on the retrieved VMR profile. For the major isotopologue, the dashed black line in panel (b) is the temperature profile. The error envelope of the minor isotopologues is quite narrow, it is within 6-8\% of the obtained isotope ratio value. The red ticks in (e) and (g) identify the $^{13}$CH$_3$CN and CH$_3$$^{13}$CN lines, respectively, in the data. The highest frequency line in (e) is the $J=30-29$, $K_a=3$ transition of C$_2$H$_5$CN (propionitrile; ethyl cyanide) at 268.0025 GHz.}

    \label{fig:model_fits_VMRs}
\end{figure*}

The minor CH$_3$CN isotopologue spectra are noisier and contain less vertical information than the major isotopologue, so we used the best-fitting CH$_3$CN main isotopologue abundance profile scaled by a uniform factor, which was varied to obtain the best fit for those species. The results of an attempted, continuously-variable CH$_3$C$^{15}$N retrieval, are discussed in Appendix \ref{sec:cont_rats}.

Finally, we tested the sensitivity of our retrieved isotopic ratios to uncertainties in the atmospheric temperature profile \citep{thelen_2020}, which vary between 1--5 K as a function of altitude. Due to the close similarities in the observed frequencies and energy levels of the different isotopologues, their line strengths scale similarly with temperature, so these uncertainties amount only to an additional error of $< 0.8\%$ on the retrieved isotopic ratios.

\section{Results}

\begin{figure}
    \centering
    \includegraphics[width=0.5\textwidth]{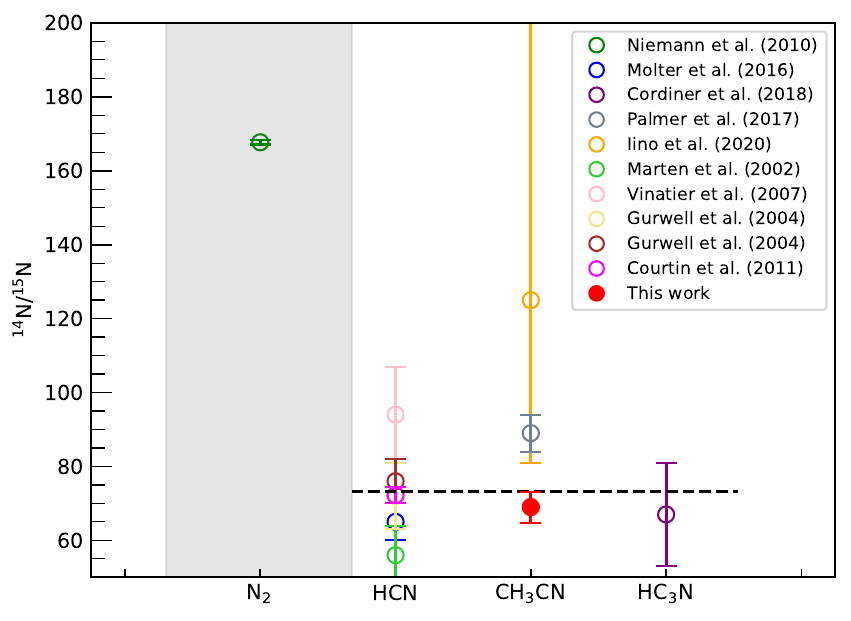}
    \caption{Previously derived $^{14}$N/$^{15}$N isotopic ratios for molecules in Titan's atmosphere, including the value derived in this work in red. The N$_2$ data point is plotted against a gray background to distinguish it as the main atmospheric nitrogen reservoir. The black dashed line is the error weighted average of previously-measured $^{14}$N/$^{15}$N ratios from \cite{iino_2020}, \cite{marten_2002}, \cite{vinatier_2007}, \cite{gurwell_2004}, \cite{courtin_2011}, \cite{molter_2016} and \cite{cordiner_2018}.}
    \label{fig:rats_comp}
\end{figure}

Our CH$_3$CN spectral fits and retrieved vertical abundance and temperature profiles are shown in Figure \ref{fig:model_fits_VMRs}. We derived scaling factors representative of the mean isotopic ratios over Titan's disk-averaged atmosphere, for the three minor CH$_3$CN isotopologues, including the $^{15}$N isotopologue and both $^{13}$C isotopologues. The CH$_3$CN/CH$_3$C$^{15}$N isotopic ratio was derived to be 68.9 $\pm$ 4.2, the CH$_3$CN/$^{13}$CH$_3$CN ratio was derived to be 89.2 $\pm$ 7.0, and the CH$_3$CN/CH$_3$$^{13}$CN ratio was derived to be 91.2 $\pm$ 6.0. Our derived $^{14}$N/$^{15}$N isotopic ratio for the major isotopologue is shown for comparison with previously obtained ratios for HCN, HC$_3$N and CH$_3$CN as well as N$_2$ in Figure \ref{fig:rats_comp}.  

Within the noise, our spectral models provide an excellent fit to the observed Titan continuum, line cores and line wings. In the lower stratosphere (below 150 km altitude), our retrieved CH$_3$CN vertical abundance (VMR) profile matches closely the steepness of the \citet{marten_2002} a-priori. Our retrieved abundances drop somewhat below the a-priori in the mid-to-upper stratosphere, with the notable exception of a maximum around 250 km, although there may be some doubt as to the physical origin of this relatively narrow feature, since it is not present in the ALMA profile retrieved by \citet{lellouch_2019}, or the photochemical model results of \citet{vuitton_2019} and \citet{dobri_loison_2018}. Considering our observations are averaged over Titan's entire Earth-facing hemisphere, the resulting spectra represent a weighted average across all latitudes and altitudes. It therefore cannot be determined where exactly (latitudinally) in Titan's atmosphere this peak originates. On the other hand, this feature is found to be necessary to obtain a good fit to the observed spectra --- otherwise, the line core and wing strengths cannot be simultaneously reproduced.

Based on the CH$_3$CN maps previously published by \citet{thelen_2019, thelen_2024} and \citet{cordiner_2019}, CH$_3$CN is most concentrated around Titan's poles. The stratospheric CH$_3$CN enhancement could be consistent with the presence of unresolved abundance peak(s) within the beam, associated with one or both of these regions. It can be speculated that the contribution is potentially coming from the trapping of molecules and subsidence in the (cold) polar regions, possibly at different altitudes around the north and south poles (see \cite{teanby_2008, teanby_2017, teanby_2019}, for example).

We then used the contribution functions from our best-fitting radiative transfer models to calculate weighted mean emission altitudes, which represent the average altitude to which our results are sensitive. These were determined to be $\sim$230 km for $^{13}$CH$_3$CN and CH$_3$C$^{15}$N (and $\sim$246 km for the major isotopologue).

\begin{figure}
    \centering
    \includegraphics[width=\columnwidth]{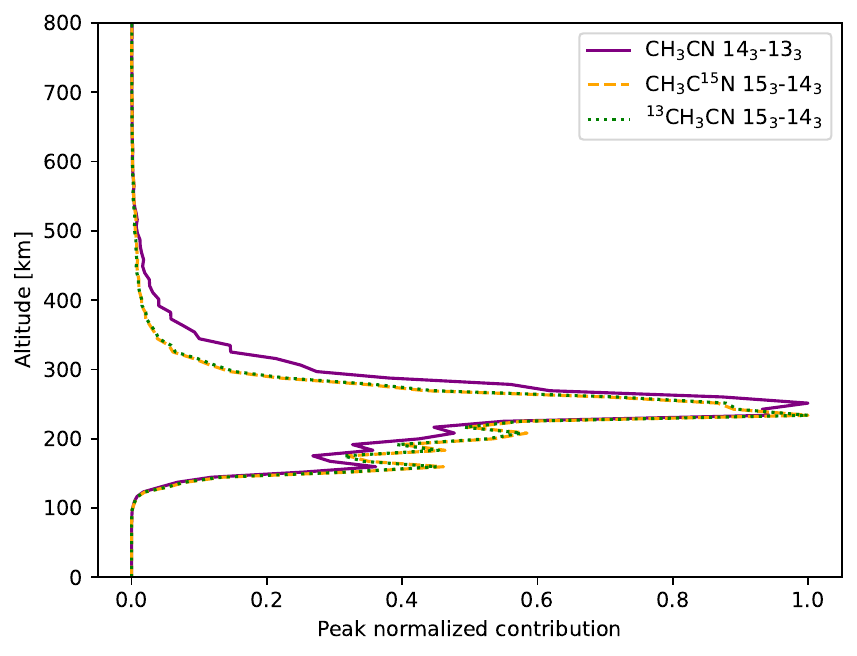}
    \caption{Contribution functions for the strongest lines ($K$=3) of CH$_3$CN and its minor isotopologues.}
    \label{fig:cont_funcs}
\end{figure}

To make a detailed comparison with the 
CH$_3$CN/CH$_3$C$^{15}$N vertical profiles produced by photochemical models, we also performed a continuously-variable CH$_3$C$^{15}$N VMR fit. However, the spectroscopic signal-to-noise ratio of 16 was found to be insufficient to provide useful constraints on the $^{14}$N/$^{15}$N ratio as a function of altitude (see Appendix \ref{sec:cont_rats}). 

\begin{figure}
    \centering
    \includegraphics[width=\columnwidth]{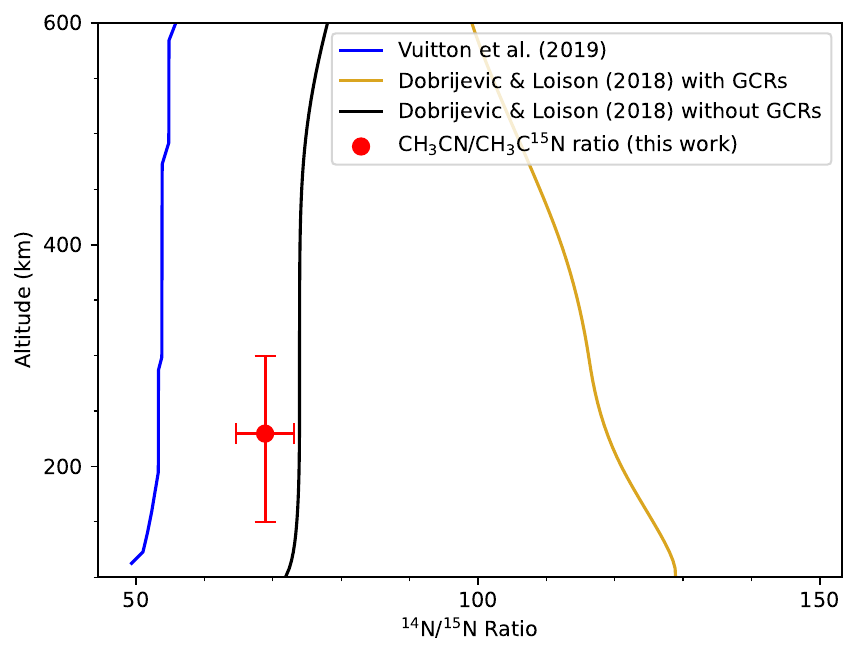}
    \caption{CH$_3$CN/CH$_3$C$^{15}$N isotopic ratio (red point) derived from our scale-factor retrieval, with horizontal error bars showing the error on the ratio value and the vertical bar spanning the region of highest sensitivity: 150-300 km. For comparison, the chemical model results of \citet{vuitton_2019} (blue line) and \citet{dobri_loison_2018} (black, without cosmic rays and gold, with cosmic rays), are shown.}
    \label{fig:photochem_models_scaled_comparison}
\end{figure}

\section{Discussion} \label{sec:discussion}

We now compare our isotopic ratios to those previously obtained in the Titan literature. We find that our CH$_3$CN/CH$_3$C$^{15}$N ratio of 68.9 $\pm$ 4.2 is significantly lower than the $^{14}$N/$^{15}$N ratio in N$_2$ of 167.7 $\pm$ 0.6 obtained by Cassini mass spectrometry (\citealt{niemann_2010}), indicating strong $^{15}$N enrichment in Titan's CH$_3$CN. On the other hand, our result is within 1.6-$\sigma$ of the previous Cassini CIRS measurement for HCN (56 $\pm$ 8; \citealt{vinatier_2007}). Additionally, our ratio is within 1-$\sigma$ of the \citet{marten_2002} HCN value of 60-70, about 2-$\sigma$ of both derived $^{14}$N/$^{15}$N ratios in HCN of 94 $\pm$ 13 and 108 $\pm$ 20 from \citet{gurwell_2004}, and is within 1.2-$\sigma$ of the \citet{courtin_2011} ratio of 76 $\pm$ 6. The HCN measurement using ALMA of 72.2 $\pm$ 2.2 by \citet{molter_2016} is also within 1-$\sigma$ of our derived ratio. Within the errors, our result is also consistent with the $^{14}$N/$^{15}$N ratio in HC$_3$N (67 $\pm$ 14; \citealt{cordiner_2018}), and within 1.3-$\sigma$ of the previous best value for the CH$_3$CN/CH$_3$C$^{15}$N ratio of 125$^{+145}_{-44}$, obtained using ALMA archival flux calibration data by \citet{iino_2020}. Although the apparent difference between our value and that of \citet{iino_2020} could be explained solely as a result of statistical noise, it should be noted that our improved values for the spectral line intensities and partition functions compared with those in the CDMS database (see Appendix \ref{sec:app_recalc_vals}) resulted in a downward revision of the $^{14}$N/$^{15}$N ratio by $\sim$18\%, $\sim$23\% for $^{12}$C/$^{13}$C and $\sim$34\% for CH$_3$CN/CH$_3$$^{13}$CN, so this could explain some of the discrepancy.

Overall, our derived ratio is in relatively good agreement with the previously obtained nitrogen isotopic ratios, with the exception of N$_2$. By far the dominant form of nitrogen in Titan's atmosphere is N$_2$, so we take the $^{14}$N/$^{15}$N ratio in N$_2$ to be representative of Titan's intrinsic isotopic composition. Isotope-selective photodissociation of N$_2$ in the upper atmosphere provides a reservoir of $^{15}$N-enriched atomic nitrogen that feeds into the altitude dependent nitrogen chemistry, resulting in the synthesis of $^{15}$N-enriched nitriles such as CH$_3$CN. In Figure 57 of \citet{vuitton_2019}, the peak density for atomic $^{14}$N and $^{15}$N occurs in Titan's thermosphere between 1100-1200 km, falling rapidly below $\sim$700 km. However, the atomic $^{14}$N/$^{15}$N isotopic ratio reaches a peak value of $\sim$25 at $\sim$900 km, below the point of the peak atomic $^{14}$N and $^{15}$N values. There are approximately constant values for the wings of the distribution of the $^{14}$N/$^{15}$N ratio: $\sim$85 at $\sim$1200 km and above and $\sim$100 at $\sim$800 km and below. Thus, the incorporation of fractionated nitrogen, and therefore production of nitriles, occurs primarily above $\sim$700 km in the \citet{vuitton_2019} model, followed by downward mixing and diffusion toward the stratosphere, into the region between 150-300 km, to which our ALMA observations are most sensitive. The enriched $^{15}$N abundance in the main nitrile production thus leads to a decrease in the $^{14}$N/$^{15}$N isotopic ratio for CH$_3$CN. The large difference between the isotopic ratios in N$_2$ and CH$_3$CN is consistent with this picture. Also comparing the $^{14}$N/$^{15}$N ratio profiles for HCN, CH$_3$CN and HC$_3$N shown in Figure 58 of \citet{vuitton_2019} reveals that they should be expected to follow a similar general trend, with decreased values at altitudes below $\sim$800 km.

Isotopic ratios for $^{12}$C/$^{13}$C have previously been measured on Titan for CH$_4$ \citep{niemann_2010}, CH$_3$D \citep{bezard_2007}, CO$_2$ \citep{nixon_2008b}, CO \citep{serigano_2016, rengel_2014, courtin_2011}, C$_4$H$_2$ \citep{jolly_2010}, C$_2$H$_2$ \citep{nixon_2008}, C$_2$H$_6$ \citep{nixon_2008}, HC$_3$N \citep{jennings_2008}, and HCN \citep{molter_2016, hidayat_1997, rengel_2014, courtin_2011}. Comparing the error weighted mean isotopic ratio for these molecules, 89.8 $\pm$ 6.9, to our derived $^{12}$C/$^{13}$C ratios in CH$_3$CN of 89.2 $\pm$ 7.0 and 91.2 $\pm$ 6.0, we find that they are all consistently within a 1-$\sigma$ error margin. The observed $^{12}$C/$^{13}$C ratios are therefore consistent with the dominant carbon reservoir in Titan's atmosphere \cite{niemann_2010,nixon_2012,mandt_2012} --- as expected considering a lack of strong isotopic fractionation mechanisms involving carbon, at the altitudes we observed. Unlike N$_2$, CH$_4$ is not subject to isotope-selective self-shielding, so variable $^{12}$C/$^{13}$C ratios among the different CH$_3$CN carbon atoms (as well as other molecules) are not expected.  Additionally, when comparing our derived $^{12}$C/$^{13}$C ratios to the error-weighted mean for comets (88.6 $\pm$ 6.5,  which excludes the outlier for H$_2$CO in comet 67P; \citealt{altwegg_2020}), and with the terrestrial value of 89.0, we also find that our measurements are within 1-$\sigma$. Thus, our $^{12}$C/$^{13}$C ratios are in good agreement with the values previously derived for Titan and various other Solar System bodies, including other bodies within the Saturnian system (see \citealt{nomura_2022}, and references therein). We thus confirm the trend for a lower degree of isotopic fractionation in carbon than for nitrogen, across various Solar System bodies.

The improved accuracy of our CH$_3$CN/CH$_3$C$^{15}$N ratio compared with \citet{iino_2020} leads to new constraints on models for the CH$_3$CN production pathways and nitrogen fractionation processes in Titan's atmosphere. The model by \citet{vuitton_2019} includes multiple pathways to CH$_3$CN, the most important being

\begin{equation}
    \mathrm{H + CH_2CN \xrightarrow{M} CH_3CN} \label{v_eq1} \\
\end{equation}
\begin{equation}  
    \mathrm{N(^2D) + C_2H_4 \longrightarrow CH_3CN + H} \label{v_eq2}\\
\end{equation}
and
\begin{equation}
    \mathrm{HCNH^+ + CH_3 \longrightarrow CH_3CNH^+} \label{v_eq3}\\
\end{equation}
followed by
\begin{equation}
    \mathrm{CH_3CNH^+ + e^- \longrightarrow CH_3CN + H} \label{v_eq4}\\
\end{equation}

\citet{balucani_2012} showed that equation \ref{v_eq2}, is inefficient since it favors the alternative isomeric product forms --- CH$_2$NCH and $c$-CH$_2$(N)CH --- rather than CH$_3$CN. \cite{dobri_loison_2018} suggested that equation \ref{v_eq1} may not be the dominant route to CH$_3$CN since it primarily occurs at the higher pressures found in the lower atmosphere.

If reaction \ref{v_eq3} (followed by reaction \ref{v_eq4}) is the primary route to CH$_3$CN (and therefore CH$_3$C$^{15}$N via HC$^{15}$NH$^+$), then this could help explain the similarity between the observed $^{14}$N/$^{15}$N isotopic ratios in HCN and CH$_3$CN.

We now compare our derived scaled nitrogen isotopic ratio with the models of \citet{vuitton_2019} and \citet{dobri_loison_2018} to gain better insight into how $^{15}$N may be incorporated into CH$_3$CN. At the value of our peak observed altitudinal sensitivity ($\approx$230 km), our value of $^{14}$N/$^{15}$N = 69$\pm$4 is in better agreement with the \citet{vuitton_2019} model value of $\sim$55 than the \citet{dobri_loison_2018} model  value of $\sim$120 (with galactic cosmic rays; GCRs included; see Figure \ref{fig:photochem_models_scaled_comparison}). On the other hand, our observed CH$_3$CN/CH$_3$C$^{15}$N ratio is a closer match with the \citet{dobri_loison_2018} model without GCRs included. However, cosmic rays are known to play a significant role in the photochemistry and ionization, primarily of Titan's lower stratosphere and troposphere \citet{nixon_2024}, so it is unclear whether this is a useful comparison. Nevertheless, it may indicate that cosmic ray chemistry should not be considered as a strong factor influencing the stratospheric CH$_3$CN/CH$_3$C$^{15}$N ratio.

The importance of magnetospheric electrons is another difference between the \citet{vuitton_2019} and \citet{dobri_loison_2018} models. In the \citet{dobri_loison_2018} model, dissociation of N$_2$ by magnetospheric electrons provides an additional source of atomic nitrogen not included in the \citet{vuitton_2019} model, particularly important in the upper atmosphere (at altitudes 700-1200 km). Since electron-impact dissociation of N$_2$ is not isotope-selective, the resulting additional source of atomic nitrogen in this altitude range has a $^{14}$N/$^{15}$N ratio equal to that of N$_2$. This causes the total atomic $^{14}$N/$^{15}$N ratio to tend towards larger values, with a corresponding impact on the $^{14}$N/$^{15}$N ratio for nitriles produced in this region. The fact that our derived isotopic ratio is greater than that of the \citet{vuitton_2019} model but less than the \citet{dobri_loison_2018} model (without GCRs), would suggest that magentospheric electrons could be important as a minor source of atomic $^{15}$N. 

While our derived isotopic ratios are in good agreement with previous measurements on Titan \citep{horst_2017}, it is of interest to compare with other Solar System bodies. Based on the data collated by \citet{nomura_2022}(and references therein), we identify some general trends for the $^{12}$C/$^{13}$C and $^{14}$N/$^{15}$N isotopic ratios. The carbon ratios are shown to have only small variations from planet to planet and most of the error bars are fairly well constrained. However, some measurements do have errors too large to constrain the exact variations. Most of the measurements for comets also maintain small variations, with the exception of the value of 40 $\pm$ 14 \citep{altwegg_2020} for H$_2$CO in comet 67P, while some have large error bars that are also unable to completely constrain the variations. Meteorites have a larger range of variation. In contrast, the nitrogen isotopic ratios are shown to generally vary amongst the objects in the solar system and may have some smaller variations amongst ratios of the same object depending on the molecule of interest. 

Given the results of our study, we can speculate about the origin of Titan's dense nitrogen reservoir. In future studies, it would be useful to gain more insight into the the amount of N$_2$ that Titan has lost since it formed; the present-day isotopic ratios of heavy, noble gases would also be useful in this regard. This could help determine if Titan's N$_2$ was formed internally through hydrothermal and cryovolcanic processes and/or if the N$_2$ was formed through photochemical reactions involving accreted cometary ices. All the nitriles observed to date show a strong $^{15}$N enrichment. As they mix downward, these nitriles condense into aerosols and precipitate onto the surface, so over time, some of the $^{15}$N is removed from the atmosphere. Therefore, the overall atmospheric $^{14}$N/$^{15}$N ratio should increase over time, suggesting that Titan's atmospheric nitrogen could have been more $^{15}$N rich in the past --- possibly more similar to the value of $\approx144$ found in comets \citep{nomura_2022}. To confirm this hypothesis will require continued, high-accuracy studies of $^{14}$N/$^{15}$N ratios in cometary nitrogen-bearing species (N$_2$, NH$_3$, HCN and other organics), as well as in the various nitrogen-bearing compounds found in Titan's surface and atmosphere.

\section{Conclusions}

Using high signal-to-noise (13-1450) ALMA observations from 2019, we derived the first well-constrained isotopologue abundance ratios for CH$_3$CN in Titan's atmosphere: 68.9 $\pm$ 4.2 for CH$_3$CN/CH$_3$C$^{15}$N, 89.2 $\pm$ 7.0 for CH$_3$CN/$^{13}$CH$_3$CN, and 91.2 $\pm$ 6.0 for CH$_3$CN/CH$_3$$^{13}$CN. These ratios represent disk-averaged values, but are most sensitive to gases in the altitude range 150-300~km, with a peak sensitivity around 230~km. We therefore show for the first time that $^{15}$N is strongly enhanced in CH$_3$CN compared to Titan's main atmospheric nitrogen reservoir (N$_2$). This can be explained as a result of photochemical isotopic fractionation initiated by isotope-selective photodissociation of N$_2$ in the thermosphere. We find a consistent level of $^{15}$N enrichment within all Titan's nitriles measured to-date, which implies they likely formed from a common, isotopically fractionated reservoir of atmospheric nitrogen. Loss of atmospheric $^{15}$N due to isotopic fractionation into nitriles should therefore be considered as a potentially important isotope sink in future studies of the time-evolution of titan's atmospheric $^{14}$N/$^{15}$N ratio, with implications for our understanding of the origins of Titan's nitrogen reservoir. Comparison between our observed nitrogen isotopic ratio and the results from previous photochemical models shows better agreement with the \citet{vuitton_2019} model than the \citet{dobri_loison_2018} that includes N$_2$ dissociation by cosmic rays and mesospheric electrons, but there is still room for improvement, so the precise importance of these processes should be revisited in future modeling efforts. Future observations at higher signal-to-noise (by at least a factor of 5 for the minor isotopologue) will be required to investigate variability in the nitrile $^{14}$N/$^{15}$N ratios as a function of altitude.

%% Also note that the akcnowlodgment environment does not support long amounts of text. If you have a lot of people and institutions to acknowledge, do not use this command. Instead, create a new \section{Acknowledgments}.
\section{acknowledgments}
The National Radio Astronomy Observatory and Green Bank Observatory are facilities of the U.S. National Science Foundation operated under cooperative agreement by Associated Universities, Inc.

This paper makes use of the following ALMA data: ADS/JAO.ALMA\#2019.1.00783.S. ALMA is a partnership of ESO (representing its member states), NSF (USA) and NINS (JAPAN), together with NRC (Canada), NSTC and ASIAA (Taiwan), and KASI (republic of Korea), in cooperation with the Republic of Chile. The Joint ALMA Observatory is operated by ESO, AUI/NRAO, and NAOJ.

The work done by MAC, AET, and CAN was funded by NASA's Solar System Observation (SSO) Program, this work was funded by the ALMA SOS program, and NAT was funded by STFC grant ST/Y000676/1. 


%\vspace{5mm}
%\facilities{HST(STIS), Swift(XRT and UVOT), AAVSO, CTIO:1.3m,


%\software{astropy \citep{2013A&A...558A..33A,2018AJ....156..123A}, 
%          Cloudy \citep{2013RMxAA..49..137F}, 
%          Source Extractor \citep{1996A&AS..117..393B}
%          }



\appendix

\section{CH$_3$CN spectroscopic parameters and partition functions} \label{sec:app_recalc_vals}

We used frequency predictions of acetonitrile (methyl cyanide; CH$_3$CN) taken from the CDMS database \citep{muller_2001} and noticed that the current CDMS predictions \citep{muller_2009} for the ground state were based on data with a significant missing window in experimental coverage around 257 GHz of current detections. The nearest experimental measurements were for the $J$=8-7 transition near 147 GHz \citep{boucher_1977} and then the $J$=18-17 transition near 331 GHz \citep{cazz_puzz_2006}. For this reason we carried out an experimental double check of the prediction accuracy in this missing window, by measuring $K$=0 to 9 transitions for $J$=14-13 at 257 GHz and $J$=15-14 at 275 GHz. Measurements were made at room temperature, and at around 1 mTorr sample pressure by using the broadband millimeter-wave (MMW) spectrometer in Warsaw \citep{medvedev_2004}. This verification turned out to be positive since CDMS predictions and currently measured frequencies for 20 different transitions were in agreement to a root mean square deviation of 16 kHz, which is well within the nominal uncertainty of the employed spectrometer.

\begin{table}[]
    \centering
    \caption{Partition functions for CH$_3$CN isotopologues as a function of temperature}
    \begin{tabular}{c c c c c}
        \hline \hline 
        T (K) & Q(CH$_3$CN) & Q($^{13}$CH$_3$CN) & Q(CH$_3$$^{13}$CN) & Q(CH$_3$C$^{15}$N) \\ [0.5ex]
        \hline
        300.0 & 10118.2635 & 10418.8811 & 10123.2613 & 10431.9556\\
        225.0 & 6570.5621 & 6765.7543 & 6573.8080 & 6774.2565 \\
        150.0 & 3576.3518 & 3682.5661 & 3578.1185 & 3687.1996 \\
        75.00 & 1265.1853 & 1302.7260 & 1265.8098 & 1304.3662 \\
        37.50 & 449.0803 & 462.3803 & 449.3016 & 462.9618 \\
        18.75 & 164.3168 & 169.1645 & 164.3975 & 169.3765 \\
        9.375 & 64.0955 & 65.9716 & 64.1267 & 66.0537 \\
        \hline
    \end{tabular}
    \label{tab:partition_fcn_vals}
\end{table}

Another issue that we faced in deriving the isotopic ratios was the need for a unified partition function for the parent species and its $^{13}$C isotopic variants. The values in the ground state entry in CDMS also accounted for the levels in the $v_8$=1 excited vibrational state, while those for the $^{13}$C species did not. For this reason we re-evaluated the partition function for the ground state of the parent at the same conditions as for the isotopic species, namely without the $v_8$=1 state, and without accounting for the nitrogen hyperfine structure, which is unresolved at the resolution of our data.  Omission of the CH$_3$CN vibrationally excited state amounts to a 7\% underestimation of the partition function. As a result of these improvements to the line intensities and partition functions, we found that the retrieved isotopic ratios decreased by 20\% for the $^{14}$N/$^{15}$N ratio, 23\% for the CH$_3$CN/$^{13}$CH$_3$CN ratio and 27\% for the CH$_3$CN/CH$_3$$^{13}$CN ratio.

\section{Line Shape Function} \label{sec:app_lsf}

In order to account for the potential effects of rotational broadening of the lines in the data, we generated a new instrumental line spread function for inclusion in NEMESIS. A rotationally broadened spectral profile was created using a modified version of the code published in \citet{cordiner_2020}, including the impact of Titan's winds on the CH$_3$CN Doppler line profile based on the observed wind field in May 2017. We convolved the rotationally broadened profile with a line-shape function that mimics the ALMA correlator response, which is the Fourier transform of a (padded) Hann window, to get the new line spread function \citep{spec_resp_memo}. Note the amount of padding used sets the frequency domain sampling and we set this to correspond to two channels, according to the standard setup of the ALMA correlator. Our observations were obtained at two different spectral resolutions (488 and 976 kHz), corresponding to a velocity FWHM of 568 and 1090 m\,s$^{-1}$ at 257.4461 and 267.5854 GHz, respectively. After the addition of wind broadening, the 488 kHz velocity FWHM increased to 604 m\,s$^{-1}$ (a difference of 36 m\,s$^{-1}$), and the 976 kHz velocity FWHM increased to 1109 m\,s$^{-1}$ (a difference of 19 m\,s$^{-1}$) so the line shape function (LSF) had a slight impact on the results.  

\begin{figure*}
\centering
    \includegraphics[width=0.45\textwidth]{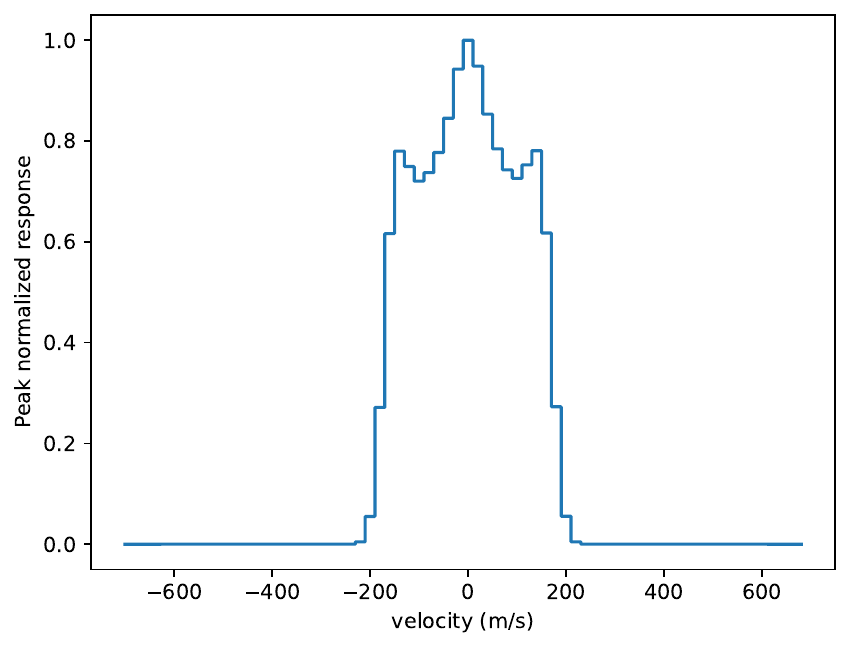}
    \includegraphics[width=0.45\textwidth]{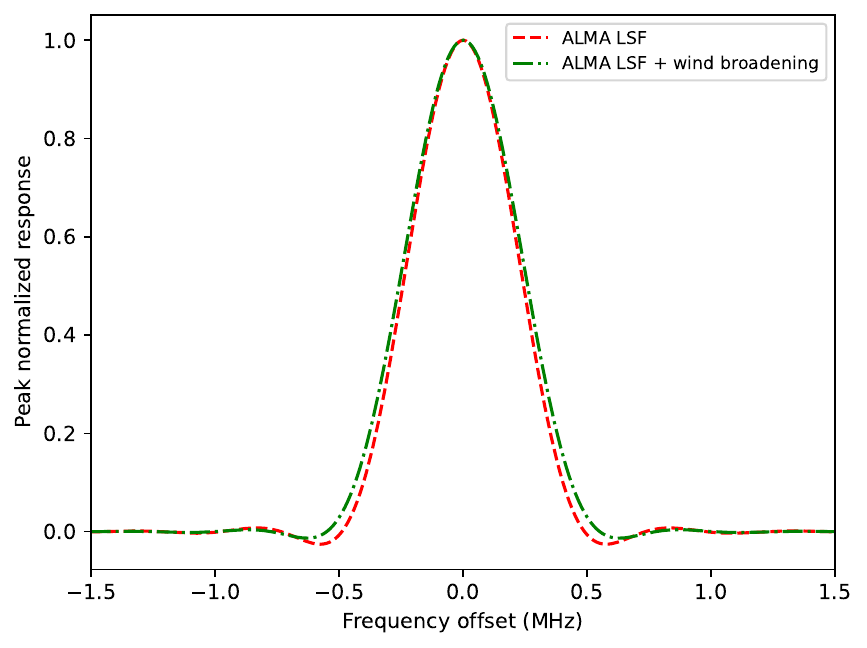}\\[3mm]
    \caption{Spectral line broadening profile based on the CH$_3$CN wind field observed by \cite{cordiner_2020} (left), and final line spread function with wind induced broadening as well as the Hanning-smoothed sinc response of the ALMA correlator (right).}
    \label{fig:rot_broad_lsf}
\end{figure*}

\section{Continuous $^{14}$N/$^{15}$N Profile Retrieval} \label{sec:cont_rats}
\begin{figure*}
    \centering
    \includegraphics[width=0.4\textwidth]{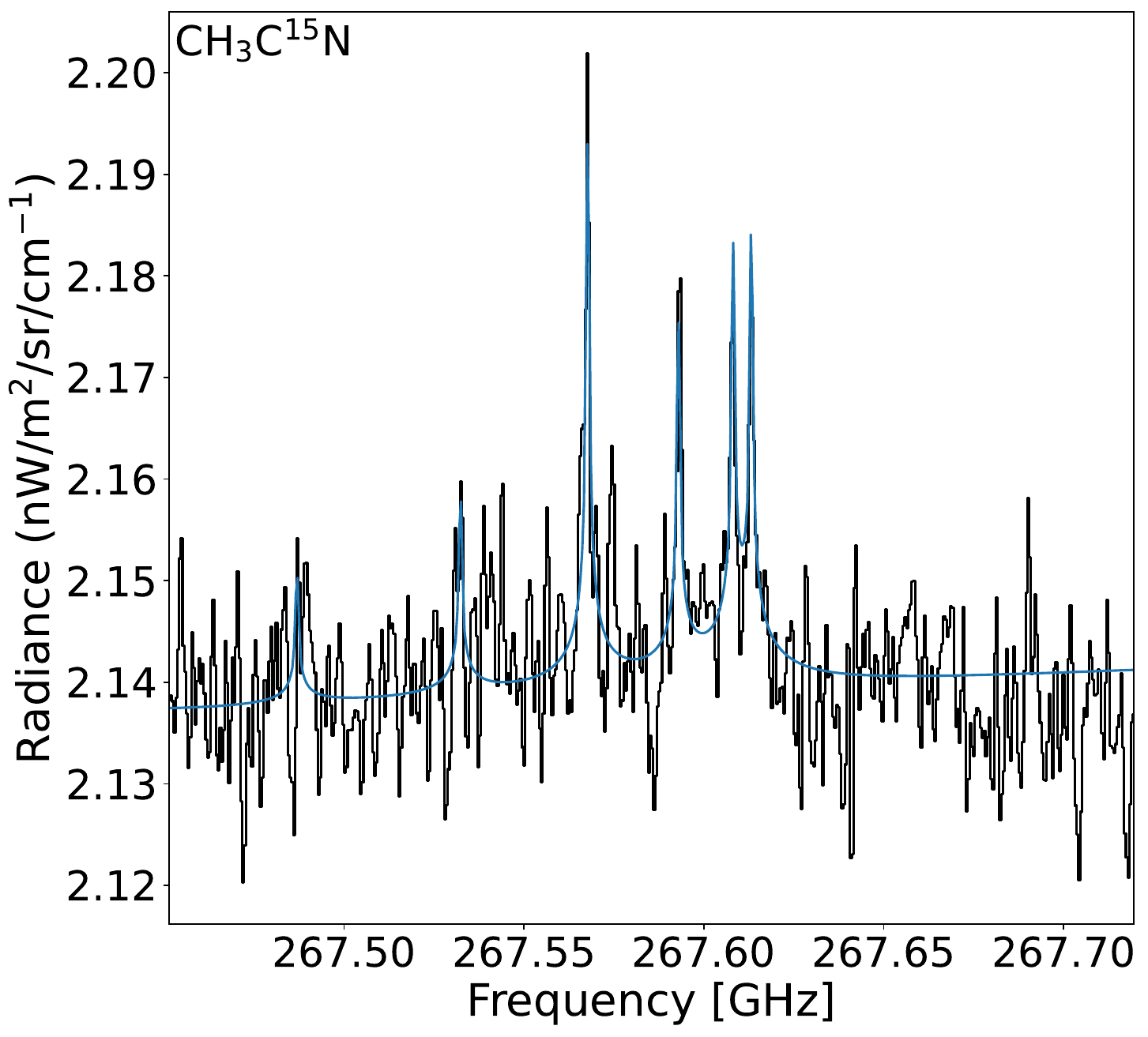}
    \includegraphics[width=0.4\textwidth]{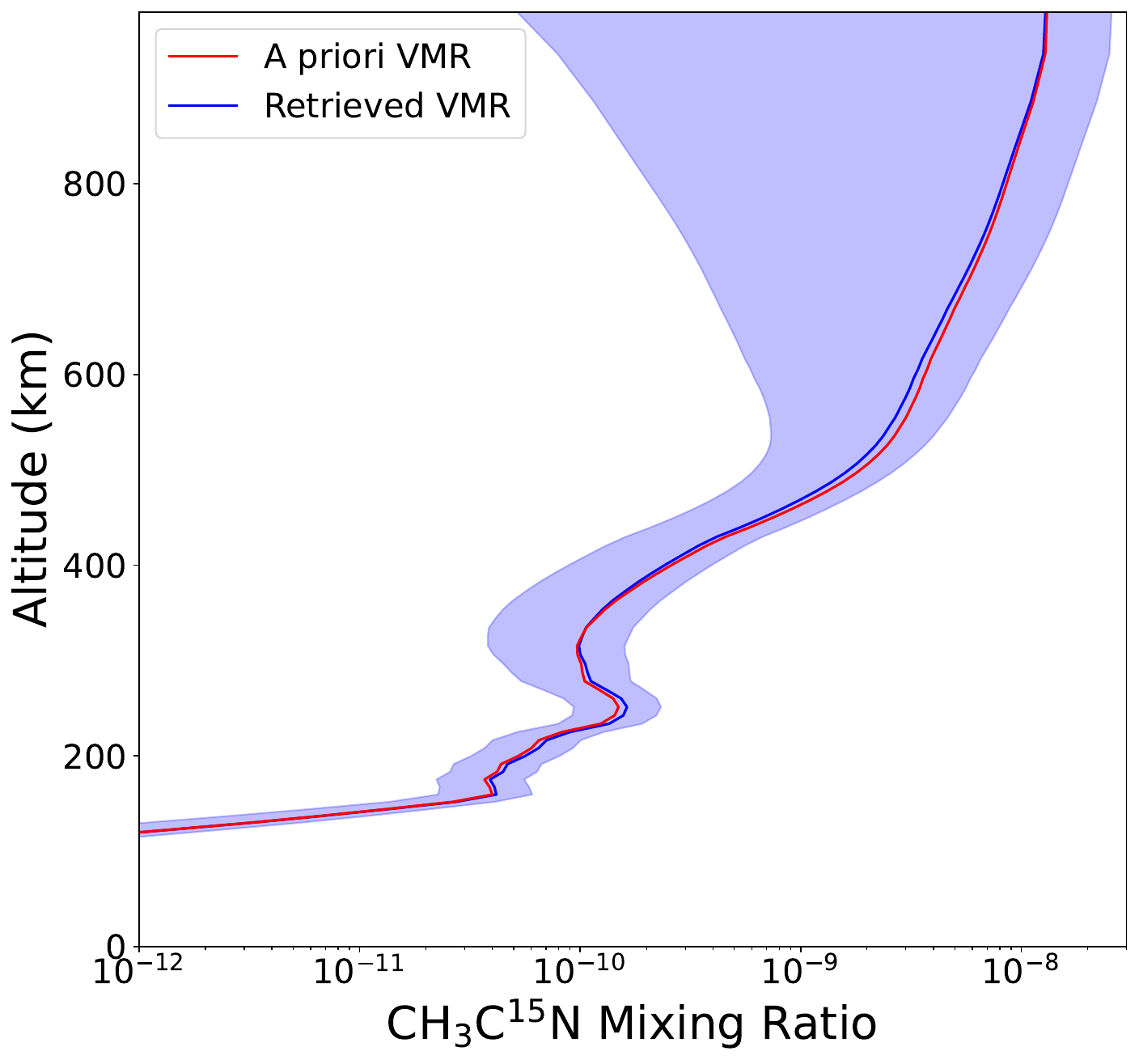}
    \caption{Continuously retrieved CH$_3$C$^{15}$N model fit (left) and VMR profile (right). As in \ref{fig:model_fits_VMRs}(a) and \ref{fig:model_fits_VMRs}(b), the data is in black and the model in blue and in the right panel the red line is the a priori profile, the blue line is the retrieved VMR profile and the shaded region is error on the retrieved VMR profile.}
    \label{fig:15N_cont}
\end{figure*}

\begin{figure}
    \centering
    \includegraphics[width=0.6\textwidth]{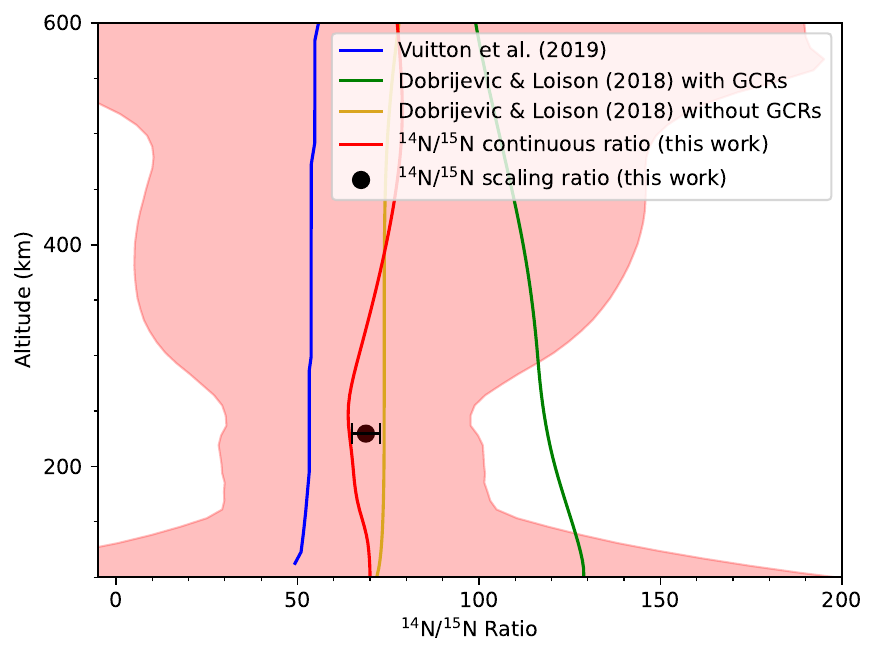}
    \caption{Derived continuous CH$_3$CN/CH$_3$C$^{15}$N isotopic ratio profile with error envelope (red; shaded red region) plotted with photochemical model profiles from \citet{vuitton_2019} (blue) and \citet{dobri_loison_2018} (green and gold). The black point is our measured scaling factor for the CH$_3$CN/CH$_3$C$^{15}$N isotopic ratio from this work, plotted at the weighted mean emission altitude of 230 km (with its corresponding error bars).}
    \label{fig:cont_rats}
\end{figure}

To investigate the vertical behavior of the CH$_3$CN/CH$_3$C$^{15}$N ratio, we used NEMESIS to perform a continuously variable abundance profile retrieval for CH$_3$C$^{15}$N, the result of which is shown in Figure \ref{fig:15N_cont}. The resulting $^{14}$N/$^{15}$N ratio and error envelope is shown in Figure \ref{fig:cont_rats}, where the errors represent the actual retrieval error, accounting for both the model and (lack of) a-priori errors. Since these errors are so large, there is no strong evidence for variability in the $^{14}$N/$^{15}$N ratio with altitude, based on our ALMA data.

\begin{figure}[p]
    \centering
    \includegraphics[width=0.6\textwidth]{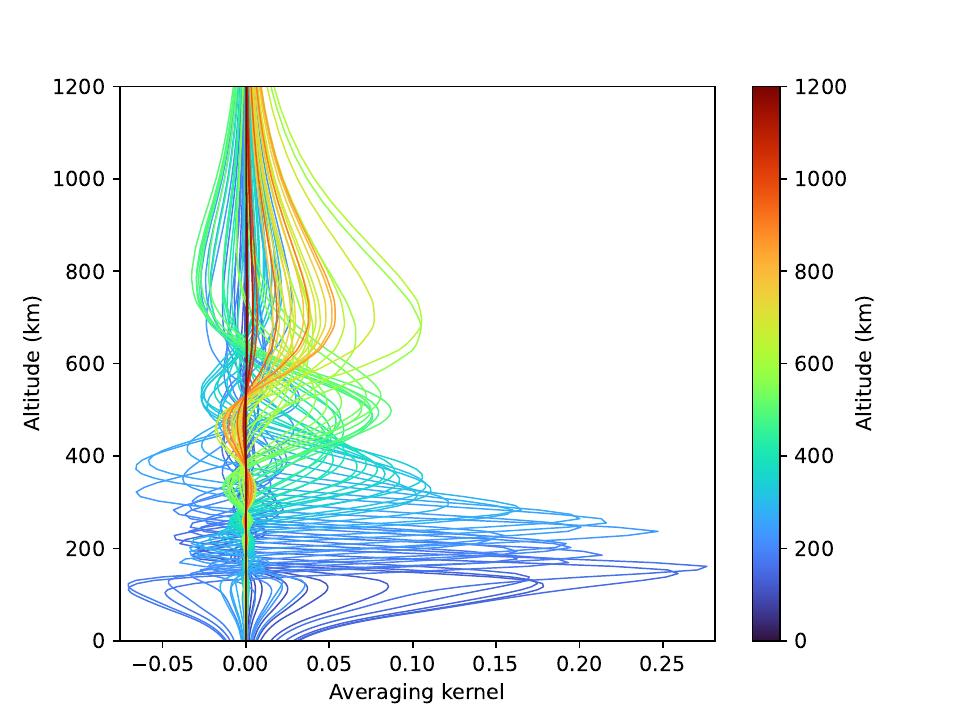}
    \caption{Averaging kernels for the CH$_3$CN major isotopologue, from our best-fitting NEMESIS model. The color bar shows the representative altitude applicable for each kernel. The altitudes with the largest values ($\sim150-300$ km) indicates the region of highest sensitivity of our model. }
    \label{fig:avg_kernels}
\end{figure}

\begin{figure}[p]
    \centering
    \includegraphics[width=0.7\textwidth]{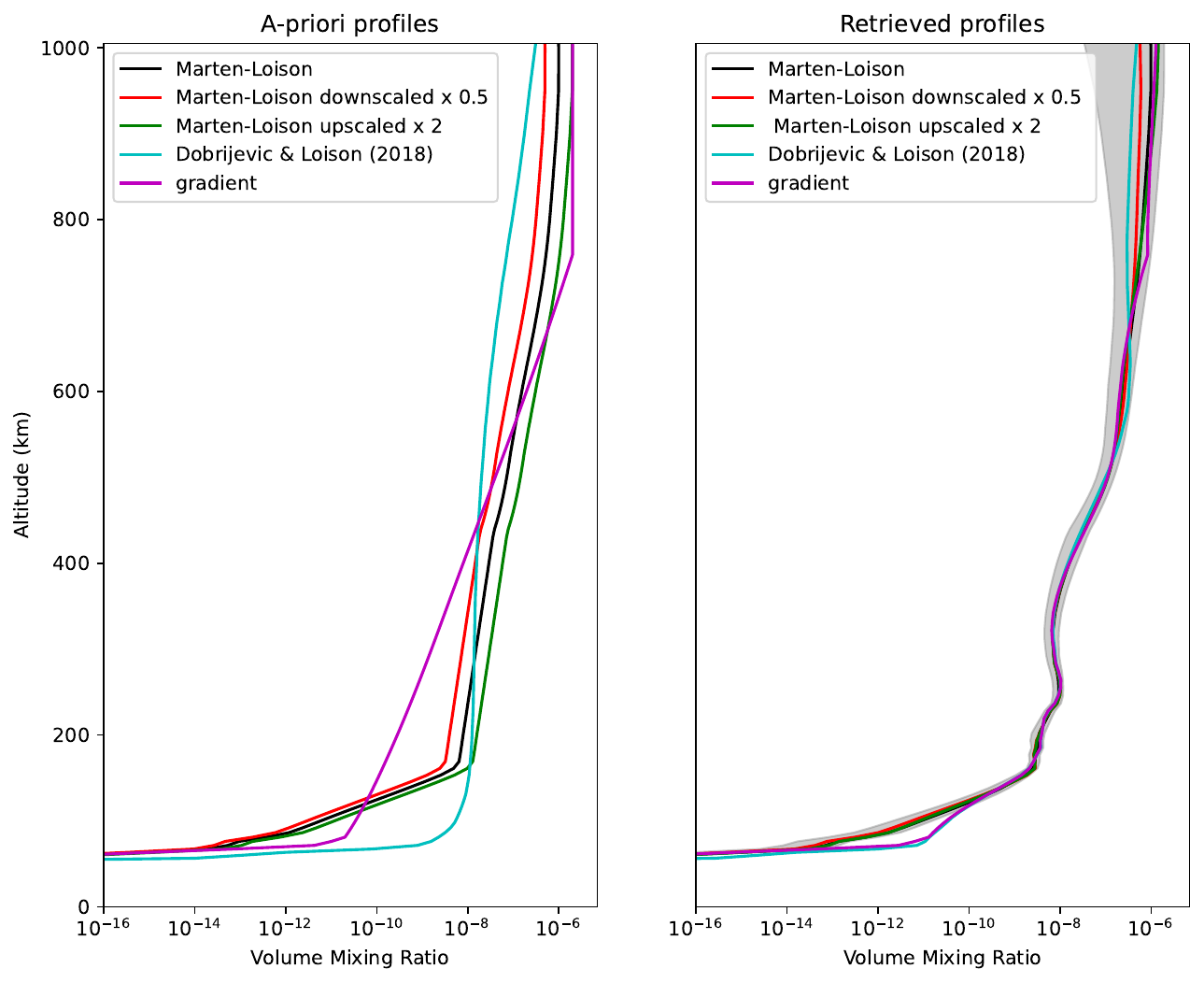}
    \caption{Plot of a-priori (left) and corresponding retrieved profiles (right) to investigate the sensitivity of the retrieved CH$_3$CN VMR to the shape of the a-priori. The retrieved CH$_3$CN VMR profile (black) is compared against retrievals using scaled versions of our default `Marten-Loison' profile (green and red profiles), the \citet{dobri_loison_2018} model profile (cyan) and a log-linear gradient profile (magenta) from \citet{thelen_2019}. The gray shaded region in the right panel is the error on the default CH$_3$CN retrieved profile.}
    \label{fig:profile_sensitivity}
\end{figure}

As discussed in section \ref{sec:discussion}, the two photochemical models make different assumptions about reaction pathways, including the weight given to interactions with magnetospheric electrons and cosmic rays, so further investigations of the $^{14}$N/$^{15}$N ratio as a function of altitude would be useful to help distinguish between the differing altitudinal dependencies as well as the impacts on the chemistry, as this will affect the abundance (and thus isotope) profiles differently at different altitudes.

\section{Averaging Kernels}
The averaging kernels for an atmospheric abundance retrieval are the product of the gain matrix ($G$) and Jacobian matrix ($J$), and provide a representation of the correlation between the retrieved abundance of each atmospheric level with respect to the other levels. Averaging kernels for our best-fitting CH$_3$CN NEMESIS model are shown in Figure \ref{fig:avg_kernels}. The averaging kernels can be used in tandem with the contribution functions to gain a better understanding of the origin of the spectral line emission and the sensitivity of the model. The altitudes with the largest kernel values (which also have the narrowest vertical envelopes), represent the region of the atmosphere to which our retrieved abundances are most sensitive, and with highest altitudinal resolution --- \emph{i.e.} in the range $\sim150$--300 km. For more details on the formal definitions of $G$, $J$, and the averaging kernel, see \citep{rodgers_1976}.

\section{Sensitivity of Retrievals to A Priori Assumptions} \label{sec:a-priori sensitivity}
Since our retrieved isotopic ratios are somewhat sensitive to the shape of the CH$_3$CN major isotopologue VMR profile, it is important to investigate the degree to which that profile may depend on any a priori model assumptions. We therefore performed a set of CH$_3$CN vertical abundance profile retrievals using different a priori profiles, including (1) scaled versions of our default a priori, (2) the CH$_3$CN profile from \citet{dobri_loison_2018}, and (3) a gradient profile from \cite{thelen_2019}.  As shown in Figure \ref{fig:profile_sensitivity}, inside the altitudinal region to which our models are sensitive ($\approx125$--500~km), the range where the peak-normalized emission contributions are greater than 1\%, we find that the retrieved CH$_3$CN abundance does not have any significant dependence on the choice of a priori.


\bibliography{sample631}{}
\bibliographystyle{aasjournal}



\end{document}